\documentclass{article}
\usepackage{icrctc07}
\title{Reconstruction accuracy of the surface detector array of the Pierre Auger Observatory}
\shorttitle{Short title}

\authors{M. Ave for Pierre Auger Collaboration$^{1}$ }
\shortauthors{Author and et al.}
\afiliations{$^1$ Pierre Auger Observatory, Av  San Mart{\'\i}n Norte 304,(5613) Malarg\"ue, Argentina}
\email{ave@cfcp.uchicago.edu}

\abstract{The reconstruction of extensive air showers (arrival
  direction, core position and energy estimation) by the surface
  detector of the Pierre Auger Observatory is discussed together with
  the corresponding accuracy.  We determine the angular reconstruction
  accuracy as a function of the station multiplicity by using two
  different aproaches. We discuss statistical and systematic
  uncertainties in the determination of the signal at 1000 m from the
  core, S(1000), which is used to estimate the primary energy. }

\begin{document}
\maketitle

\section{Introduction}
The Pierre Auger Observatory consists of two independent components: 
the fluorescence detector (FD) and the surface detector (SD) 
\cite{auger}. We have determined the angular resolution of events 
recorded by the surface detector alone, on an event by event basis, 
from the zenith ($\theta$) and azimuth ($\phi$) uncertainties
obtained from the geometrical reconstruction, using the relation 
described in \cite{icrc2005}:
~$F(\eta)~=~1/2~(V[\theta]~+~\sin^2({\theta})~V[\phi])~$, where 
$\eta$ is the space-angle, and $V[\theta]$ and $V[\phi]$ are the
variance of $\theta$ and $\phi$ respectively. We define the angular 
resolution ($AR$) as the angular radius that would contain 68\% of
showers coming from a point source, 
$AR~=~1.5~\sqrt{F(\eta)}$. We checked the angular resolution using the 
redundant information given by a sub-array composed by adjacent detectors.



The parameter used to infer the energy of the surface detector events
($S(1000)$) is studied and its systematic and statistical errors are 
determined. The event-by-event error estimation is checked with full
Monte Carlo simulations . The unavoidable fluctuations
in this parameter caused by fluctuations in the shower development is
evaluated with simulations for different primary assumptions.

\section[Angular Resolution]{Angular Resolution}\label{angreso} 

The arrival direction of a SD event is determined by fitting the arrival
time of the first particle in each station to a shower front model.
The precision achieved in the arrival direction depends, on the clock
precision of the detector and on the fluctuations in the first
particle arrival time. In \cite{timevariance} an empirical model
has been developed to determine the uncertainty in the  
time measurement of each individual detector participating in the
event.

The model of the shower front used in the minimization procedure, 
be it spherical, parabolic, or even planar also influences the
uncertainty in the arrival direction determination, but not as 
much as the time measurement precision. It has been shown in 
\cite{timevariance} that a parabolic model for the shower front 
adequately describes the data.

\vspace{0.2cm}
{\bf \large{On a event by event basis}}

\begin{figure}
\begin{center}
\noindent
\vspace{-1.8cm}
\includegraphics
    [width=0.49\textwidth,height=0.47\textwidth]{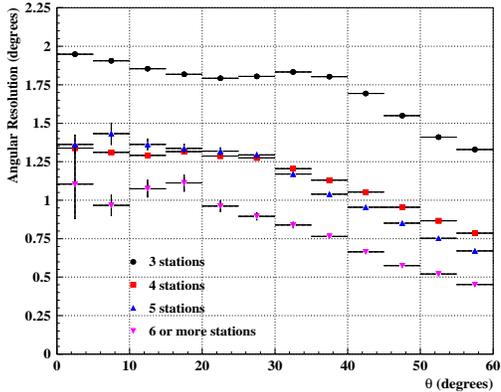}
\vspace{-1.1cm}
\end{center}
\caption{Angular resolution ($AR$) for the
	  SD as a function of the zenith angle ($\theta$). 
	  The $AR$ is plotted for various station multiplicities. 
} \label{fig:prof} 
\end{figure}

Given the two inputs: a model for the time variance and a model for 
the shower front, the angular resolution can be calculated on an event
by event basis out of a minimization procedure. In Fig.~\ref{fig:prof}, 
we show our angular resolution as a function of the zenith angle for 
various station 
multiplicities (circles: 3 stations, squares: 4 stations, up
triangles: 5 stations, down triangles: 6 stations or more).
The data used to build this plot spans from January/2004 to March/2007. 


As it can be seen, the angular resolution is better than $2^\circ$
in the worst case of vertical showers with only 3 stations hit. This
value improves significantly for 4 or 5 stations\footnote{For 4 and 5 
stations the AR is very similar because in the fitting procedure they 
have the same number of degrees of freedom.}. 
For 6 or more stations, which corresponds to events with energies
above 10 EeV, the angular resolution is in all cases better than
about~$1^\circ$. 

\vspace{0.2cm}
{\bf \large { Using station pairs }}

\begin{figure}

\begin{center}
\vspace{-0.3cm}
\noindent
\includegraphics
    [width=0.44\textwidth]{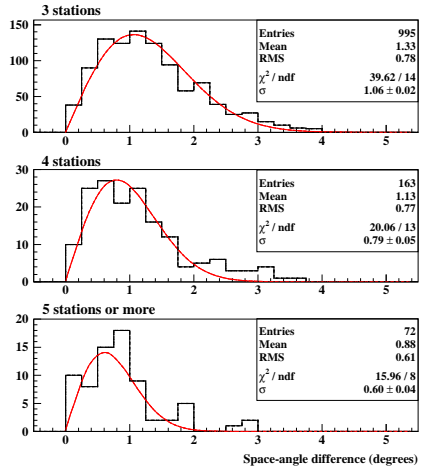}
\vspace{-0.7cm}
\end{center}
\caption{Space-angle difference between two SD estimates of the event 
 geometry for different multiplicities (see text for more details).}
\label{fig:twins}
\end{figure}

A new sub-array of pairs has been recently deployed as a part of
the Surface Detector array. These are adjacent detectors located $\sim$~11~m
apart, and therefore are sampling the same region of the shower
front. To do this analysis, events with at least three pairs are
selected. The reconstruction is then performed twice, each
time using the time information of one of the tanks in each pair. This 
provide two quasi-independent estimates of the
geometry. In Fig.~\ref{fig:twins} we show the space-angle difference
between these two estimates for showers with 3, 4, and 5 or
more stations.

The distributions are then fitted to the adjusted Gaussian resolution 
function
($dp \propto e^{-\eta^2/2\sigma^2}~d($cos$(\eta))~d\eta$, where $\eta$ is the 
angle between the two reconstructions of the same shower) to
obtain $\sigma$. The angular resolution (68\% contour), which is
given by 1.5 times $\sigma$, is in agreement with the one obtained 
on a event by event basis.

\section[Energy Estimator]{Energy Estimator}\label{s1000reso} 

\begin{figure}
  \begin{center}
    \noindent
    \vspace{-0.5cm}
    \includegraphics[width=0.39\textwidth]{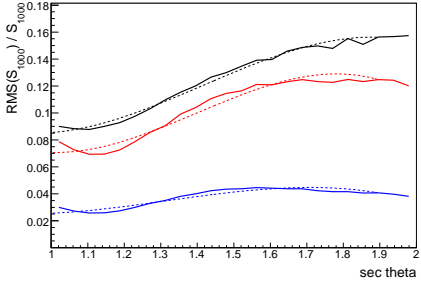}
    \vspace{-0.1cm}    
  \end{center}
  \caption{Relative spread due to shower to shower fluctuations 
    for different compositions (blue-iron, red-proton, black-mixed 
    composition).} \label{fig:shsh}
\end{figure}

The surface detector only samples the properties of an air shower at a
limited number of points at different distances from the shower axis
($r$).  An observable has to be then defined to estimate the shower
size. To avoid the large fluctuations in the signal integrated over
all distances caused by fluctuations in the shower development,
Hillas \cite{Hillas} proposed to use the signal at a given distance
($S(r)$) to classify the size of the shower. In Fig.~\ref{fig:shsh} we
show the predictions from Monte Carlo simulations of the magnitude
of the fluctuations in $S(r=1000)$ as a function of zenith angle. The
relative fluctuations are found to be independent of energy and its 
magnitude is $\sim$ 10\% for most of the cases studied.

The experimental error in the estimation of the signal size at a given 
core distance depends  on the spacing of the array. 
In \cite{Newton} it has been shown that for the Auger array spacing
the optimum distance ($r_{opt}$) to minimize this experimental error
is $\sim$1000~m. Therefore, the observable that we use to relate to
the primary energy will be the signal size at 1000~m
($S(1000)$)\footnote{$S(1000)$ is measured in units of VEM, i.e. the
  signal produced by a vertical centered muon.}. However, it should be 
noted that $r_{opt}$ fluctuates 
from event to event and increases to larger core distances ($\sim$ 
1500~m) when there are saturated stations \cite{Newton}.

To estimate $S(1000)$ it is necessary to adopt a lateral distribution
function (LDF) that describes the fall-off of the signal size with
the distance to the shower axis. The function used here is a modified
NKG function given by:
$S(r)=S(1000)\left(\frac{r}{1000}\right)^{-\beta}\left(\frac{r+700}{1700}\right)^{-\beta}$,
where $r$ is the distance to the shower axis in meters, $S(r)$ is the
signal size at a core distance $r$, $S(1000)$ is the size parameter
of the shower, and $\beta$ is called the slope of the LDF.

\vspace{0.2cm}
{\bf \large {S(1000) uncertainties}}

The signal sizes in each station are then used to estimate the core
location and $S(1000)$, with $\beta$ being a fixed parameter. The
$fitting$ error in $S(1000)$ is a consequence of the uncertainty of
the observed signal size largely due to the finite dimension of the
detectors. This will be the statistical error in $S(1000)$
($\sigma^{stat}_{S(1000)}$). The uncertainty in the signal sizes has
been measured directly using pairs of stations located close to each
other in the field \cite{signalaccuracy}.

\begin{figure}
  \begin{center}
    \vspace{-0.6cm}
    \includegraphics[width=0.45\textwidth]{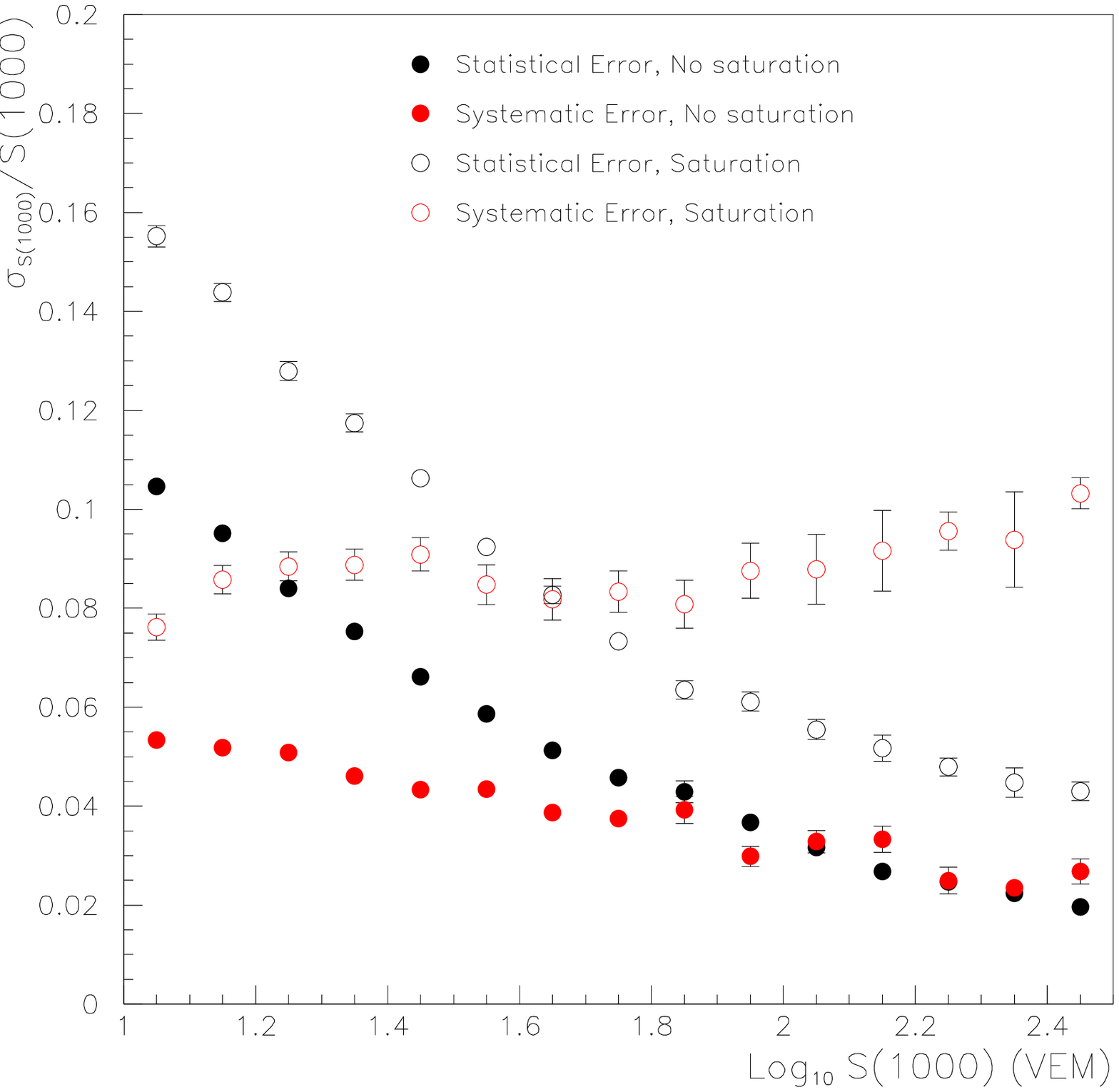}
    \vspace{-0.8cm}
  \end{center}
  \caption{The average systematic and statistical error in S(1000) 
    as a function of $log~S(1000)$. 
    The data has been divided in two sets (events with-without 
    stations saturated). }
  \label{fig:s1000sys-stat}
\end{figure}

The second source of error in $S(1000)$ is a systematic
 ($\sigma^{sys}_{S(1000)}$) arising from the lack of knowledge of the
 true LDF shape for a particular event.  If
 the $r_{opt}$ of a given event is close to 1000 m, the fitted
 $S(1000)$ is independent of the value of $\beta$ assumed
 \cite{Newton}.  When it is not, fluctuations in the event by event
 $\beta$ give rise to a systematic error. The value of $\beta$ to be
 used in the reconstruction has been estimated empirically: in a small
 subset of events ($S(1000)>$ 20 VEM and having more than 5 stations) the
 $\beta$ is left as a free parameter as well. We then parameterize the
 fitted values of $\beta$ as a function of zenith angle and $S(1000)$. The deviation
 from this parameterization is calculated for each event and the RMS
 ($\sigma_\beta$) parameterized as a function of $S(1000)$ (no
 dependence on zenith angle has been found). The result is the
 following:~$\sigma_\beta(S(1000))= 0.71 \times \exp(
 -0.976~\log(S(1000)))$.
We then repeat $N$ times the reconstruction of each event, fixing $\beta$ to values
 sampled from a Gaussian distribution centered around the prediction
 with the sigma given above. The RMS of the fitted $S(1000)$ from
 these set of fits is then the systematic error of $S(1000)$
 ($\sigma^{sys}_{S(1000)}$).

In Fig. \ref{fig:s1000sys-stat} we show the average
 systematic and statistical error of $S(1000)$ as a
 function of $\log(S(1000))$. The data has been divided in two sets: events
 with (without) saturated stations. Two features are clearly seen: a)
 the error in events with saturated stations is systematically 5\% larger,
 b) $\sigma^{stat}_{S(1000)}$ dominates the error budget for
 $S(1000)< $40~VEM. No dependence of $\sigma^{sys}_{S(1000)}$ or
 $\sigma^{stat}_{S(1000)}$ on zenith angle has been found.


\begin{figure}
  \begin{center}
    \vspace{-0.6cm}
    \noindent
    \includegraphics[width=0.43\textwidth]{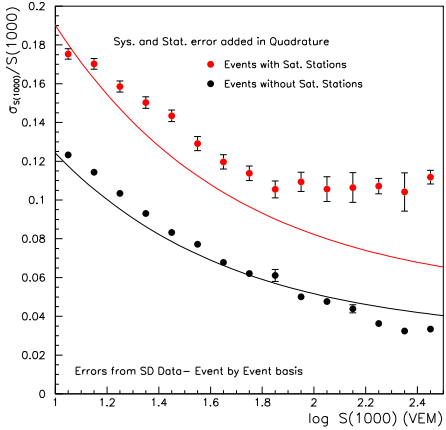}
    \vspace{-0.8cm}
  \end{center}
  \caption{Total error in $S(1000)$ calculated on an event
    by event basis from the data. The data is separated in two sets: events
    with (without) saturated stations. The lines correspond to the predictions
    from full MC calculations (see text for details)}
  \label{fig:toterr-comp}
\end{figure}

\vspace{0.2cm}
{\bf \large{Using Full Monte Carlo Simulations}}

To benchmark our error estimation we have created a library of Corsika
 showers for proton primaries, zenith angles $\theta$=
 0-12-25-36-45-60~degrees 
and energies
 $\log_{10}E(eV)$=17.8-18.0-18.2-18.4-18.6-19.0-19.5-20.0. For each
 Corsika shower, we calculate the $true$ $S(1000)$ and it is then used 
 to generate 10 (25) events (depending on the energy) with random core
 positions. 



The reconstruction procedure used for the data is then applied to the simulations.
For each zenith angle and energy we fit the distribution of $\log
 \left(\frac{S(1000)^{rec}}{S(1000)^{true}}\right)$ to a Gaussian
 function. The mean value and sigma are then parameterized as a function
 of $S(1000)^{true}$. No zenith angle dependence has been found. A bias in the
 reconstructed $S(1000)$ is only found for $S(1000)<$ 10 VEM.  The
 sigma of this distribution is the quadrature combination of the
 statistical and systematic error in $S(1000)$. In Fig.~\ref{fig:toterr-comp}
we show the comparison of the sigma of these distributions with the
 average total error obtained on an event by event basis. The data
 is separated in two sets: events with (without) saturated stations.
 The circles correspond to the total error obtained on a event by event
 basis, the lines are the predictions from full Monte Carlo simulations.
The agreement is excellent except for a slight overestimation of the
 error ($\sim$4\%) for saturated events at large energies.

\section{Conclusions} 

The angular resolution of the surface detector was determined experimentally, 
 checked using the pairs data set and  found to be better than 
2$^\circ$ for 3-fold events ($E<4$~EeV), better than 1.2$^\circ$ for  
4-folds and 5-folds events ($3<E<10$~EeV) and better than 0.9$^\circ$ 
for higher multiplicity events ($E>10$ ~EeV).

The error of the parameter used to infer the energy of the surface
 detector events ($S(1000)$) has been determined experimentally, 
checked using full Monte Carlo simulations 
and found
 to be  better than 8\%
 (12\%) at the highest energies for events with (without) saturated
 stations.  At high energies, the fluctuations in $S(1000)$ are dominated by
 fluctuations in the shower development.


\end{document}